\documentclass[letterpaper,11pt,onecolumn]{IEEEtran}

\IEEEoverridecommandlockouts

\usepackage{graphicx}
\usepackage{psfrag}

\usepackage[cmex10]{amsmath}
\usepackage{amssymb}
\usepackage{amsfonts}

\usepackage{cases}
\usepackage{units}
\usepackage{dsfont}

\usepackage{cite}

\newcommand{\E}{\mathrm{E}}

\DeclareMathOperator{\dB}{dB}
\DeclareMathOperator{\tr}{\mathbf{tr}}
\DeclareMathOperator{\diag}{\mathbf{diag}}

\providecommand{\abs}[1]{\lvert{#1}\rvert}

\title{Efficient Linear Precoding in Downlink Cooperative Cellular Networks with Soft Interference Nulling}

\author{%
	\authorblockN{Chris~T.~K.~Ng and Howard Huang}\\
	\authorblockA{Bell Laboratories, Alcatel-Lucent, Holmdel, NJ 07733\\
		Email: \{Chris.Ng, hchuang\}@alcatel-lucent.com}
}

\begin{document}

\maketitle
\thispagestyle{empty}

\begin{abstract}

A simple line network model is proposed to study the downlink cellular network. Without base station cooperation, the system is interference-limited. The interference limitation is overcome when the base stations are allowed to jointly encode the user signals, but the capacity-achieving dirty paper coding scheme can be too complex for practical implementation. A new linear precoding technique called soft interference nulling (SIN) is proposed, which performs at least as well as zero-forcing (ZF) beamforming under full network coordination. Unlike ZF, SIN allows the possibility of but over-penalizes interference. The SIN precoder is computed by solving a convex optimization problem, and the formulation is extended to multiple-antenna channels. SIN can be applied when only a limited number of base stations cooperate; it is shown that SIN under partial network coordination can outperform full network coordination ZF at moderate SNRs.

\end{abstract}


\section{Introduction}

Interference management is a fundamental challenge in wireless cellular systems.
In this paper, we consider the downlink cellular network, and investigate the performance benefits of allowing cooperation and joint processing among the base stations.
Without base station cooperation, the system is interference-limited, i.e., the signal-to-interference-plus-noise ratio (SINR) at the mobiles cannot be improved simply by increasing the base station transmit power, since higher transmit power also creates stronger interference.
Given the deployment of a fixed number of base stations, one approach to increase system throughput is to allow the joint encoding of user signals across the base stations.
In this case, assuming perfect cooperation among the base stations, the downlink system can be modeled as a broadcast channel (BC).
However, the theoretically optimal dirty paper coding (DPC) transmission scheme for the BC can be too complex for practical implementation.
Zero-forcing (ZF) beamforming is a simple linear precoding technique that offers good performance in a BC\@.
In this paper, we propose a new linear precoding technique called soft interference nulling (SIN) that performs better than or equal to ZF\@.
The SIN precoder can be found by solving a convex optimization problem.
Moreover, we show that SIN can be applied when the terminals have multiple antennas, as well as in the case when each user is served by overlapping coordination clusters each with only a limited number of cooperating base stations.

The time division multiplexing access (TDMA), ZF, and DPC rates in downlink cellular networks are compared in \cite{viswanathan03:dl_cap_cell_itfr, sharif07:ts_dpc_bf_mimo_bc_users}, and the performance of ZF is studied in \cite{sharif05:mimo_bc_part_si, yoo06:opt_ma_bc_zf_bf}.
Different precoding schemes for multiple-input multiple-output (MIMO) BCs are presented in \cite{caire03:ach_guas_mimo_bc, peel05:vec_pert_mimo_inv_reg, hochwald05:vec_pert_mimo_perturbation}.
The optimality of DPC in a MIMO BC is shown in \cite{weingarten06:cap_mimo_bc}.
For single-cell multiuser MIMO channels, the optimization of different performance metrics in terms of the user rates or SINRs are considered in
\cite{kobayashi06:iter_wf_wtsr_mimo_bc, stojnic06:rate_max_bc_lin_proc, christensen09:wt_srate_mmse_mimo_bc, wiesel06:lin_pre_conic_opt_mimo, shi07:dl_mmse_opt_mu_dual, shi08:dl_mmse_opt_mmse_bal, shi08:rate_opt_mu_mimo_lin_proc, zhang08:mimo_bc_mac_cov_constr}. In this paper, we consider the maximization of a general concave utility function of the user rates under the assumption that interference is treated as noise.
Cooperating base stations for the cellular uplink channel is considered in \cite{venkatesan07:coop_uplink_spectral, venkatesan07:coop_uplink_prop_fair}.
Capacity gain from transmitter and receiver cooperation is investigated in \cite{ng07:cap_tx_rx_coop}.
When the user signals are jointly encoded by separate base stations, they are under per-antenna power constraints (PAPC).
ZF under PAPC are considered in \cite{boccardi06:zf_mimo_bc_papc, karakayali07:opt_zf_papc, wiesel08:zf_pre_gen_inv},
and DPC under PAPC is treated in \cite{yu07:tx_opt_ma_dl_papc}.

The remainder of this paper is organized as follows.
The channel system model and capacity bounds are described in Section~\ref{sec:sys_mod}.
Section~\ref{sec:coop_base_sta} considers cooperative base stations and zero-forcing (ZF) beamforming.
Section~\ref{sec:soft_int_null} presents the soft interference nulling (SIN) precoding technique under different coordination cluster sizes, with an extension to multiple-antenna channels.
Numerical results are presented in Section~\ref{sec:num_res},
and Section~\ref{sec:conclu} concludes the paper.

\paragraph*{Notation}
In this paper, $\mathds{R}_+^N$ is the set of $N$-dimensional nonnegative real vectors; $\mathds{H}_+^N$ is the set of $N\times N$ positive semidefinite Hermitian matrices; $\mathds{C}$ denotes the complex field; $[A]_{i,j}$ is the $(i,j)$ entry of the matrix $A$; $\det$, $\tr$ denote determinant and trace, respectively; and $\diag(a)$ is a diagonal matrix with its diagonal given by the vector $a$.


\section{System Model}
\label{sec:sys_mod}

\subsection{Channel Model}

\begin{figure}
  \centering
  \psfrag{P}[][]{$P$}
  \psfrag{N}[][]{$N$}
  \psfrag{dx}[][]{$d_x$}
  \psfrag{dy}[][]{$d_y$}
  \includegraphics{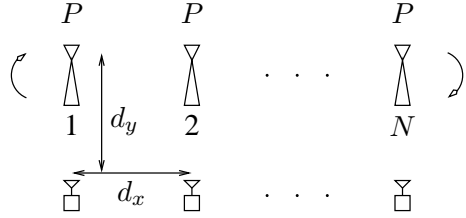}
  \caption{A line network of base stations.}
  \label{fig:base_lnet}
\end{figure}

We consider a simple model of the wireless cellular downlink network.
Suppose there are $N$ base stations in the network, and they are positioned along a line with distance $d_x$ apart.
Each base station serves one mobile user, and we assume each user is located a distance $d_y$ away from its base station, as illustrated in Fig.~\ref{fig:base_lnet}.
To minimize the boundary effects, we consider a line network with wraparound where the distance between User~$i$ and Base~$j$ is given by
\begin{align}
d_{ij} &\triangleq \sqrt{d_y^2+\bigl(d_xd(i,j)\bigr)^2}, \quad i,j = 1,\dotsc,N
\end{align}
where
\begin{align}
d(i,j) &\triangleq \min(\lvert i-j \rvert, \lvert i-j+N \rvert, \lvert i-j-N \rvert).
\end{align}

We consider a narrow-band flat-fading channel model.
Wireless systems with wider bandwidth may be modeled as multiple narrow-band channels using modulation schemes such as orthogonal frequency-division multiplexing (OFDM), and most techniques discussed in this paper remain applicable.
For now let us assume each terminal is equipped with a single omnidirectional antenna; multiple antenna base stations and mobiles are considered in Section~\ref{sec:mimo_ch}.
Suppose $x_j \in \mathds{C}$ is the transmit signal at Base~$j$, and
$y_i \in \mathds{C}$ is the receive signal at User~$i$.
The discrete-time channel model is then described by
\begin{align}
y_i &= \sum_{j=1}^N h_{ij} x_j + z_i,\quad i=1,\dotsc,M
\end{align}
where $h_{ij} \in \mathds{C}$ is the complex baseband channel,
and $z_i \sim \mathcal{CN}(0,1) \in \mathds{C}$ is zero-mean circularly symmetric complex Gaussian (ZMCSCG) noise normalized with unit variance.
For each User~$i$, the desired signal is $h_{ii}x_i$, and the inter-cell interference from other base stations is given by $\sum_{j=1,j\neq i}^N h_{ij}x_j$.
Therefore, a frequency reuse factor of $1$ is assumed.
Alternatively, under other frequency reuse patterns, the system model may represent the group of base stations that occupy the same frequency band.
The radio signal propagation from any base station to any user is modeled as independent Rayleigh fading with a power attenuation factor proportional to $d^\eta$, where $d$ is the propagation distance, and $\eta$ is the path loss exponent:
i.e., each entry of $h_{ij}$ is independent and identically distributed (i.i.d.) as $\mathcal{CN}(0,d_{ij}^{-\eta})$.
We assume $\eta=4$, which corresponds to the path loss in a typical outdoor urban cellular environment.
Note that in the line network system model, each mobile user suffers from two dominant interferers, which is similar to the case as in a three-sector hexagonal cellular network.

We consider a block-fading channel model:
the channels realize independently according to their distribution at the beginning of each fading block, and they remain unchanged within the duration of the fading block.
In this paper, we assume the channel states can be estimated accurately and conveyed timely to all base stations: i.e., the channels are known at all terminals.
Each base station is under a transmit power constraint of $P$.
We consider a short-term power constraint: i.e., $\E[\abs{x_j}^2] \leq P$, $j=1,\dotsc,M$, where the expectation is over repeated channel uses within a fading block; power allocation across fading blocks is not considered.
We assume each fading block is sufficiently long so the transmitters may code at channel capacity using random Gaussian codewords.

\subsection{Achievable rates and capacity bounds}
\label{sec:rates_bc_bounds}

In traditional cellular systems, the base stations do not cooperate.
Assuming each base station transmits at full power,
then User~$i$ receives at the rate
\begin{align}
R_{i,\text{Int}} = \log\biggl( 1+\frac{\lvert h_{ii}\rvert^2 P_i}
{1+\sum_{j\neq i} \lvert h_{ij} \rvert^2 P_j} \biggr).
\end{align}
In this paper, we wish to investigate efficient transmission schemes that exploit base station cooperation and joint processing.
A performance upper bound can be obtained by considering the capacity of a cellular network with perfect base station cooperation.
Suppose each base station knows the messages of all users, and we allow joint encoding at the base stations.
Then this cooperative cellular system may be modeled as a broadcast channel (BC) with $N$ single-antenna receivers, and an $N$-antenna transmitter under per-antenna power constraints (PAPC).
For a Gaussian multiple-input multiple-output (MIMO) BC, its capacity region \cite{weingarten06:cap_mimo_bc} is achieved by the dirty paper coding (DPC) scheme \cite{costa83:writing_dirty_paper}.
In DPC, the messages for the users are encoded in a given order, and the interference from the previously encoded users are pre-subtracted at the transmitter for the subsequently encoded users.
We consider the sum rate of the MIMO BC as a performance metric for the cooperative cellular system.
In \cite{yu07:tx_opt_ma_dl_papc}, it is shown that the sum rate of a MIMO BC under PAPC can be found by solving the following convex minimax optimization problem
\begin{align}
\min_{q}\; \max_{s} \quad &\log \det \Bigl(\sum_{i=1}^{N}s_i\tilde{h}_i\tilde{h}_i^H + \diag(q)\Bigr) - \sum_{i=1}^{N} \log q_i\\
\text{over} \quad & q\in \mathds{R}_+^N,\; s\in \mathds{R}_+^N\\
\text{subject to} \quad
& \sum_{i=1}^N s_i \leq NP\\
& \sum_{i=1}^N q_i \leq N
\end{align}
where
\begin{align}
\label{eq:htil_q_s}
\tilde{h}_i &\triangleq [h_{i1} \dots h_{iN}]^T \in \mathds{C}^N, &
q &\triangleq [q_1 \dots q_N]^T \in \mathds{R}_+^N, & s&\triangleq [s_1 \dots s_N]^T \in \mathds{R}_+^N.
\end{align}

\begin{figure}
  \centering
  \includegraphics*[width=8cm]{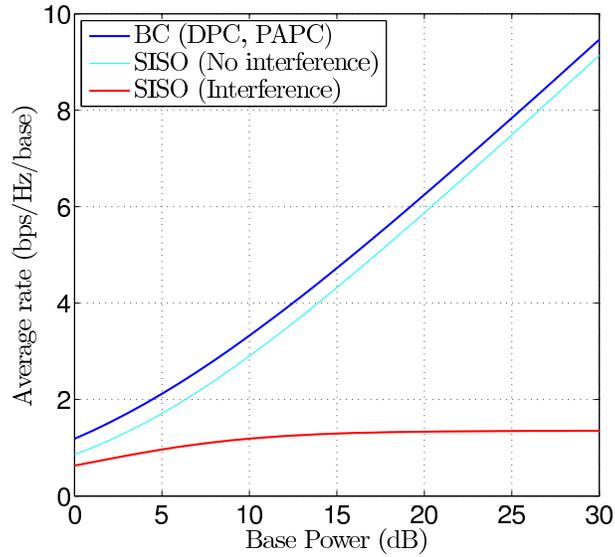}
  \caption{Single-user rates (with and without interference) and BC capacity in the downlink line network ($N=19$, $d_x=d_y=1$).}
  \label{fig:lnet_rates_P_N19_BC_SISO}
\end{figure}

The achievable rate $R_{i,\text{Int}}$ and the BC sum rate are shown in Fig.~\ref{fig:lnet_rates_P_N19_BC_SISO} (solid lines) as a function of the SNR $P$ for a cellular line network with $N = 19$ base stations, and geometry $d_x = d_y = 1$.
In Fig.~\ref{fig:lnet_rates_P_N19_BC_SISO}, $50$ sets of random channel realizations are generated.
For each set of channel realizations, the non-cooperative and the cooperative sum rates are calculated.
Then the rates are averaged over the random channel realizations, and normalized by the number of base stations.
The normalized per-base non-cooperative and cooperative rates are not particularly sensitive to the number of base stations $N$, which justifies the consideration of a wraparound line network model of moderate size.
For comparison, also shown in Fig.~\ref{fig:lnet_rates_P_N19_BC_SISO} (dotted line) is the single-input single-output (SISO) rate in the absence of interference
\begin{align}
R_{i,\text{No-Int}} = \log(1+\lvert h_{ii}\rvert^2 P_i).
\end{align}

Without base station cooperation, at increasing SNR $P$, the average SINR at the mobile saturates at approximately $2\,\dB$.
It matches well with the typical operating SINR of current cellular systems, which justifies the choice of $d_x,d_y$.
Henceforth, we assume $d_x = d_y = 1$.
Fig.~\ref{fig:lnet_rates_P_N19_BC_SISO} illustrates the interference-limited nature of cellular systems: the user rates under interference fail to keep increasing with SNR as in the case when interference is absent.
On the other hand, the interference limitation can be overcome by allowing the base stations to cooperate, as shown by the DPC rates when the cooperative system is modeled as a BC.\@
However, as the complexity of DPC can be challenging for practical implementation, in this paper we explore efficient linear precoding techniques that achieve near the DPC performance.

\section{Cooperative Base Stations}
\label{sec:coop_base_sta}

In this section, let us consider the case where all base stations participate in the joint encoding of the messages of all users.
Partial network coordination is considered in Section~\ref{sec:part_coord}.
When the base stations cooperate, it is convenient to consider the transmit signals of all base stations jointly as a vector.
For notational convenience, we define
\begin{align}
x &\triangleq [x_1 \dots x_N]^T \in\mathds{C}^N, & y &\triangleq [y_1 \dots y_N]^T \in\mathds{C}^N, & z &\triangleq [z_1 \dots z_N]^T \in\mathds{C}^N.
\end{align}
Hence the channel written in matrix form is
\begin{align}
y &= Hx+z, & H &\triangleq [\tilde{h}_1 \dots \tilde{h}_N]^T \in\mathds{C}^{N\times N}
\end{align}
where $x$ is encoded jointly by all base stations, but each $y_i$ is decoded separately by User~$i$.

\subsection{Zero-Forcing Beamforming}

A simple linear transmit precoding technique is zero-forcing (ZF) beamforming \cite{caire03:ach_guas_mimo_bc}.
Unlike DPC, no interference pre-subtraction is performed at the transmitter.
Instead, the transmitter chooses a set of precoding beamforming weights such that interference is zeroed out at each mobile user.
Suppose $u\in\mathds{C}^N$ denotes the information signals for the $N$ users, with $\E[uu^H] = I_N$.
Then the ZF beamforming transmit signal is given by
\begin{align}
x &= W\diag(a)u
\end{align}
where $W\in\mathds{C}^{N\times N}$ is the beamforming precoding matrix,
and $a\triangleq [a_1 \dots a_N]^T \in\mathds{R}_+^N$ controls the effective channel gains of the users.
Since we have the same number of users as the number of transmit antennas, assuming $H$ is full rank, we set $W = H^{-1}$ to zero out interference at each user.
With such choice of $W$, the downlink system decouples into a set of $N$ interference-free parallel channels
\begin{align}
y_i &= a_iu_i + z_i, \quad i = 1,\dotsc,N.
\end{align}
Define the effective channel power gain for User~$i$ as $\gamma_i \triangleq a_i^2$.
Zero-forcing in MIMO BC subject to PAPC is considered in \cite{boccardi06:zf_mimo_bc_papc, karakayali07:opt_zf_papc, wiesel08:zf_pre_gen_inv}.
In particular, the ZF sum rate under PAPC can be found by solving the following convex optimization problem
\begin{align}
\text{maximize}\quad & \sum_{i=1}^N \log(1+\gamma_i)\\
\text{over}\quad & \gamma \in \mathds{R}_+^N\\
\text{subject to}\quad &
\label{eq:max_ZF_W_P1}
\lvert W \rvert^2 \gamma \leq P\mathbf{1}
\end{align}
where $\gamma \triangleq [\gamma_1 \dots \gamma_N]^T$, $W = H^{-1}$, and $\mathbf{1}$ denotes a vector of $1$'s.
The last constraint in (\ref{eq:max_ZF_W_P1}) represents component-wise inequality, and
$\lvert W \rvert^2$ denotes the component-wise squared magnitude of the entries of $W$: i.e., $\lvert W \rvert^2 \triangleq [\lvert w_{ij}\rvert^2]$.
In general, ZF is suboptimal. However, at high SNR with multiuser diversity, ZF performs asymptotically close to the optimal DPC scheme \cite{sharif05:mimo_bc_part_si, yoo06:opt_ma_bc_zf_bf, sharif07:ts_dpc_bf_mimo_bc_users}.
Zero-forcing requires all base stations to cooperate in order to ensure an interference-free signal at each of the users.
In the next section, we consider a linear precoding technique that is applicable under partial network coordination, where only a limited number of base stations cooperate, and the cooperating base stations may form overlapping clusters.

\section{Soft Interference Nulling}
\label{sec:soft_int_null}

\subsection{Full Network Coordination}

First we consider the case where all base stations are cooperating. Partial network coordination is addressed in the next section.
Let the joint transmit signal at the base stations be given as follows:
\begin{align}
x &= \sum_{i=1}^N G_i u_i, \quad i = 1,\dotsc,N
\end{align}
where $G_i\in\mathds{C}^{N\times N}$ is the precoding matrix for User~$i$, and $u_i\in\mathds{C}^N$ is User~$i$'s information signals.
Note that in the above formulation we allow a user's transmission to have multiple spatial streams (up to the number of transmit antennas $N$).
Without loss of generality, entries of a user's precoding matrix may be set to zero if not all spatial streams are active.
We assume each spatial stream consists of independent data signals, i.e., $\E[u_iu_i^H] = I_N$, and $\E[u_iu_j^H] = \mathbf{0},\, i\neq j$.
Treating interference as noise, the transmission rate to User~$i$ is given by
\begin{align}
R_i &= \log \biggl(1+\frac{\tilde{h}_i^H Q_i \tilde{h}_i}{1+\sum_{j=1,j\neq i}^N\tilde{h}_i^H Q_j \tilde{h}_i}\biggr), \quad i=1,\dotsc,N
\end{align}
where
$\tilde{h}_i$ are as defined in (\ref{eq:htil_q_s}), and
$Q_i \in \mathds{H}_+^N$ is the covariance matrix of the transmit signal to User~$i$
\begin{align}
Q_i &= \E[G_iu_i(G_iu_i)^H] = G_iG_i^H.
\end{align}
It is more convenient to specify the precoder design in terms of the covariance matrices $Q_i$.
To generate transmit signals with the specified covariances, we may set the precoding matrices $G_i$ to be
\begin{align}
G_i = V_iD_i^{\nicefrac{1}{2}}, \quad i=1,\dotsc,N
\end{align}
where $D_i$ is diagonal, and $Q_i = V_iD_iV_i^H$ is the eigendecomposition of the covariance matrix $Q_i$.
Suppose we consider as performance metric a general concave utility function of the user rates: $U(R)$, where $R \triangleq [R_1 \dots R_N]^T$, then the design goal is to choose the covariance matrices $Q_i$ to maximize the utility function
\begin{align}
\label{eq:max_U_R}
\text{maximize} \quad & U(R)\\
\text{over} \quad & R\in\mathds{R}_+^N,\; Q_i \in \mathds{H}_+^N\\
\text{subject to} \quad
& R_i \leq \log \biggl(1+\frac{\tilde{h}_i^H Q_i \tilde{h}_i}{1+\sum_{j=1,j\neq i}^N\tilde{h}_i^H Q_j \tilde{h}_i}\biggr)\\
\label{eq:max_U_R_Pj}
& \Bigl[\sum_{j=1}^N Q_j\Bigr]_{i,i} \leq P
\end{align}
where $i=1,\dotsc,N$, and the last set of inequalities represents the per-antenna power constraints.
However, the above maximization is not a convex program, and in general it is difficult to solve efficiently.
Instead, we propose a linear precoding technique called soft interference nulling (SIN),
which has good performance in the sense that SIN precoding performs at least as well as ZF beamforming under full network coordination.

The main idea is that when the interference at each user is small, the transmission rates $R_i$ are well-approximated by the following:
\begin{align}
R_i &= \log \biggl(1+\frac{\tilde{h}_i^H Q_i \tilde{h}_i}{1+\sum_{j=1,\,j\neq i}^N\tilde{h}_i^H Q_j \tilde{h}_i}\biggr)\\
&= \log \Bigl(1+\sum_{j=1}^N\tilde{h}_i^H Q_j \tilde{h}_i\Bigr) - \log \Bigl(1+\sum_{j=1,\,j\neq i}^N\tilde{h}_i^H Q_j \tilde{h}_i\Bigr)\\
\label{eq:Ri_gts_log_tr_siso}
&\gtrsim \log \Bigl(1+\sum_{j=1}^N\tilde{h}_i^H Q_j \tilde{h}_i\Bigr) - \sum_{j=1,\,j\neq i}^N\tilde{h}_i^H Q_j \tilde{h}_i\\
&\triangleq \tilde{R}_i
\end{align}
where the inequality in (\ref{eq:Ri_gts_log_tr_siso}) follows from $x$ being a global over-estimator of $\log(1+x)$, and $\log(1+x)\approx x$ for small $x$.
We define the SIN precoders as the precoding matrices that correspond to the solution of following optimization problem
\begin{align}
\text{maximize} \quad & U(\tilde{R})\\
\text{over} \quad & \tilde{R}\in\mathds{R}_+^N,\; Q_i \in \mathds{H}_+^N\\
\text{subject to} \quad
& \tilde{R}_i \leq \log \Bigl(1+\sum_{j=1}^N\tilde{h}_i^H Q_j \tilde{h}_i\Bigr) - \sum_{j=1,\,j\neq i}^N\tilde{h}_i^H Q_j \tilde{h}_i\\
& \Bigl[\sum_{j=1}^N Q_j\Bigr]_{i,i} \leq P,
\end{align}
where $i=1,\dotsc,N$, and $\tilde{R} \triangleq [\tilde{R}_1 \dots \tilde{R}_N]^T$.
Note that the above maximization is a convex optimization problem, and its solution can be efficiently computed using standard convex optimization numerical techniques, e.g., by the interior point method \cite{renegar01:math_ipm_cvxopt, boyd04:convex_opt}.

When the interference terms are zero, the inequality in (\ref{eq:Ri_gts_log_tr_siso}) is tight, and the SIN rates are the same as the ZF rates.
Unlike ZF, however, SIN allows the possibility of nonzero interference:
ZF can be interpreted as imposing an infinite penalty on interference, whereas SIN relaxes such restriction.
Nevertheless, as given in (\ref{eq:Ri_gts_log_tr_siso}), SIN over-penalizes interference (i.e., $\tilde{R}_i \lesssim R_i$);
hence there is a strong incentive for the SIN precoding solution to null out the interference.
Since ZF beamforming is in the feasible set of the SIN optimization, SIN performs at least as well as ZF\@.
Note that the SIN solution is not necessarily locally optimal in the original problem (\ref{eq:max_U_R})--(\ref{eq:max_U_R_Pj}): the utility function may be further improved, for example, by a local gradient search method.
However, when there is sufficient transmit antenna degrees-of-freedom to suppress the interference, such improvements are observed to be marginal in the numerical experiments.

\subsection{Partial Network Coordination}
\label{sec:part_coord}

Having full network coordination implies that all user messages need to be routed to all base stations, which represents a considerable burden on the backhaul network.
In practice, it is desirable to involve only a limited number of base stations cooperating to send to each user.
These cooperating base stations form a cluster, and we may have overlapping coordination clusters in the cellular network.
In this section, we consider applying SIN precoding under partial network coordination.
In particular, we assume the base stations may exchange channel state information (CSI) over the backhaul network, but each user's message is routed only to those base stations in the coordination cluster associated with the user.

Suppose the set of base stations that cooperate to send to User~$i$ is given by the ordered set $J_i = \{j_1,j_2,\dotsc,j_{N_i}\}$,
where $j_1 < j_2 < \dotsb < j_{N_i}$.
Hence only $N_i$ base stations, with $N_i \leq N$, cooperate to send to User~$i$, and we impose the constraint that base stations not in $J_i$ do not participate in the transmission of User~$i$'s signals.
For each User~$i$, the above restriction is represented by an association matrix $C_i$ defined as follows:
\begin{align}
C_i &\in \mathds{R}^{N\times N_i}, &
[C_i]_{j_l,l} &=
\begin{cases}
1, & l=1,\dotsc,N_i\\
0, & \text{otherwise}.
\end{cases}
\end{align}
Suppose the joint transmit signal for User~$i$ among its $N_i$ cooperating base stations is characterized by the $N_i\times N_i$ covariance matrix: $Q_i \in \mathds{H}_+^{N_i}$,
then the full $N\times N$ covariance matrix for User~$i$ with respect to all base stations is given by: $C_iQ_iC_i^T$.
Effectively, the matrix $C_i$ specifies that the non-participating base stations have precoding weights of zero.
The SIN precoding covariance matrices under partial base station coordination can then be found by solving the following convex optimization problem
\begin{align}
\text{maximize} \quad & U(\tilde{R})\\
\text{over} \quad & \tilde{R}\in\mathds{R}_+^N,\; Q_i \in \mathds{H}_+^{N_i}\\
\text{subject to} \quad
& \tilde{R}_i \leq \log \Bigl(1+\sum_{j=1}^N\tilde{h}_i^H C_jQ_jC_j^T \tilde{h}_i\Bigr) - \sum_{j=1,\,j\neq i}^N\tilde{h}_i^H C_jQ_jC_j^T \tilde{h}_i\\
& \Bigl[\sum_{j=1}^N C_jQ_jC_j^T\Bigr]_{i,i} \leq P
\end{align}
where $i=1,\dotsc,N$.

For example, suppose there are $N=5$ base stations in the network, and in the network User~$3$ specifies that its coordination cluster consists of Bases~$2$, $3$, and $4$:
i.e., Bases $2$, $3$, and $4$ may cooperate and jointly encode the signals to be sent to User~$3$, but Bases~$1$ and $2$ may not participate to send any message to User~$3$.
Then the the association matrix for User~$3$ is given by
\begin{align}
C_3 = \begin{bmatrix}
0 & 0 & 0\\
1 & 0 & 0\\
0 & 1 & 0\\
0 & 0 & 1\\
0 & 0 & 0
\end{bmatrix}.
\end{align}
The design variable for User~$3$'s signal is given by the covariance matrix $Q_3 \in \mathds{H}_+^3$, which describes the joint signal from Bases~$2$, $3$, and $4$.
With respect to all base stations, the overall covariance matrix for User~$3$'s signal is described by
\begin{align}
C_3Q_3C_3^T = \begin{bmatrix}
0 & 0 & 0 & 0 & 0\\
0 & [Q_3]_{1,1} & [Q_3]_{1,2} & [Q_3]_{1,3} & 0\\
0 & [Q_3]_{2,1} & [Q_3]_{2,2} & [Q_3]_{2,3} & 0\\
0 & [Q_3]_{3,1} & [Q_3]_{3,2} & [Q_3]_{3,3} & 0\\
0 & 0 & 0 & 0 & 0
\end{bmatrix}.
\end{align}
Numerical examples with limited coordination cluster size are presented in Section~\ref{sec:num_res}.

\subsection{Multiple-Antenna Channels}
\label{sec:mimo_ch}

In the previous sections, we assume each mobile user has a single antenna.
The SIN precoding technique generalizes to the case where the users have multiple antennas.
To illustrate the principle, we consider a simple single-cell scenario.
Suppose we have a base station with $M_T$ transmit antennas, and there are $N$ mobile users in the network each with $M_R$ receive antennas.
We assume there is a transmit power constraint $P$ on the base station.

The multiple-input multiple-output (MIMO) downlink channel is described by
\begin{align}
y_i &= H_i x + z_i,\quad i=1,\dotsc,N
\end{align}
where $y_i\in\mathds{C}^{M_R}$ is User~$i$'s receive signal; $x\in\mathds{C}^{M_T}$ is the transmit signal; $z_i\in\mathds{C}^{M_R}$ is additive white Gaussian noise (AWGN) with $\E[z_iz_i^H] = I_{M_R}$, $\E[z_iz_j^H] = \mathbf{0}$ for $i\neq j$; and
$H_i\in\mathds{C}^{M_R\times M_T}$ is the MIMO channel from the base station to User~$i$.
Based on the Taylor series expansion of the log-determinant function: $\log\det(I+X) = \tr X + O(X^2)$, the user rates are approximated as
\begin{align}
R_i &= \log \frac{\det\bigl(I_{M_R} + \sum_{j=1}^N H_iQ_jH_i^H\bigr)}{\det\bigl(I_{M_R} + \sum_{j=1,\,j\neq i}^N H_iQ_jH_i^H\bigr)}\\
&\gtrsim \log \det\Bigl(I_{M_R} + \sum_{j=1}^N H_iQ_jH_i^H\Bigr) - \sum_{j=1,\,j\neq i}^N \tr H_iQ_jH_i^H\\
&\triangleq \tilde{R}_i.
\end{align}
Therefore, the MIMO SIN precoding covariance matrices can be found by solving the following convex optimization problem
\begin{align}
\text{maximize} \quad & U(\tilde{R})\\
\text{over} \quad & \tilde{R}\in\mathds{R}_+^N,\; Q_i \in \mathds{H}_+^{M_T}\\
\text{subject to} \quad
& \tilde{R}_i \leq \log \det\Bigl(I_{M_R} + \sum_{j=1}^N H_iQ_jH_i^H\Bigr) - \sum_{j=1,\,j\neq i}^N \tr H_iQ_jH_i^H\\
& \sum_{j=1}^N \tr Q_j \leq P
\end{align}
where $i=1,\dotsc,N$.
Note that unlike ZF, SIN precoding is well-defined even when the number of transmit antennas is less than the number of users.
However, since SIN imposes a large penalty on interference, we expect the SIN formulation is most useful when there are sufficient degrees-of-freedom to suppress the interference among the users.

\section{Numerical Results}
\label{sec:num_res}

\begin{figure}
  \centering
  \includegraphics*[width=12cm]{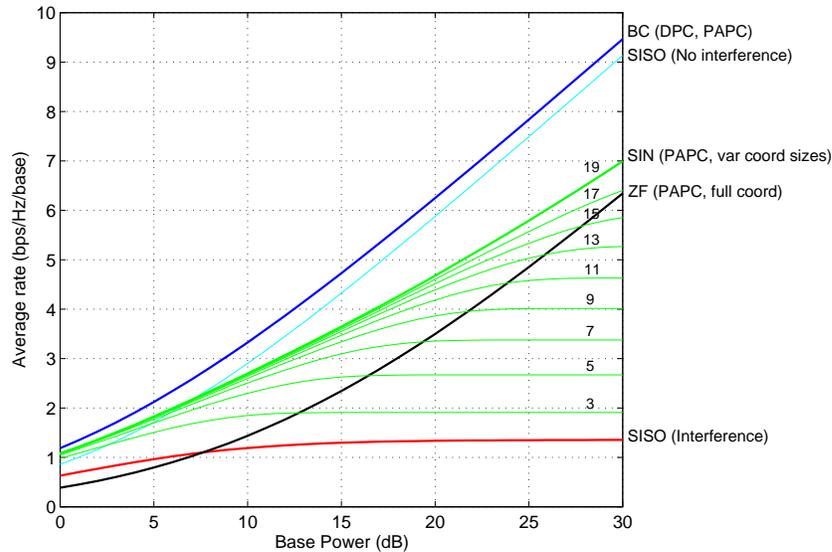}
  \caption{ZF and SIN (with different coordination cluster sizes) rates in the downlink line network ($N=19$, $d_x=d_y=1$).
  The SIN cluster sizes are labeled next to their corresponding curves.}
  \label{fig:sin_co_lnet_Nc_P_N19_Nr50_2}
\end{figure}

Let us consider the line network cellular downlink model and the simulation settings as described in Section~\ref{sec:sys_mod}.
The rates obtained from ZF and SIN precoding are shown in Fig.~\ref{fig:sin_co_lnet_Nc_P_N19_Nr50_2}.
For the numerical results, the convex optimization problems are solved using the software package SDPT3 \cite{tutuncu03:sdpt3}.
For comparison, the single-user rates (with and without interference) and the DPC BC capacity discussed in Section~\ref{sec:rates_bc_bounds} are also shown on the plot.
There are $N=19$ single-antenna base stations in the network, and each base station serves one single-antenna user.
In ZF, all base stations participate in the cooperation.
For SIN precoding, different coordination cluster sizes are considered, and the cluster sizes are labeled next to their corresponding curves on the plot.
Each user chooses the closest base stations to participate in the coordination cluster.
For example, when the coordination cluster size is $3$, User~$1$ is served by Bases~$\{19,1,2\}$, User~$2$ by Bases~$\{1,2,3\}$, User~$3$ by Bases~$\{2,3,4\}$, and so on.

As illustrated in Fig.~\ref{fig:sin_co_lnet_Nc_P_N19_Nr50_2}, there is a gap between the ZF rate and the DPC rate, but they both exhibit similar scaling trends as the SNR $P$ increases.
In particular, the ZF beamforming technique allows the network to overcome its interference-limited performance bottleneck, thus demonstrating the value of cooperative cellular networks.
Under full network coordination, SIN precoding outperforms ZF, especially at low SNRs where ZF can suffer from noise amplification.
Moreover, at moderate SNRs, SIN with limited coordination cluster sizes is able to outperform ZF, which requires full network coordination.
For example, at the SNR $P = 18\,\dB$, a typical operating SNR at the cell edge in cellular systems, SIN with a coordination cluster size of $7$ outperforms ZF with full network coordination.
As the SNR increases, however, it is observed that SIN under partial network coordination becomes interference-limited.
It is useful to identify the region to the lower-right of the ZF rate curve as the interference-limited regime.
Therefore, for power efficiency, the coordination cluster size in SIN could be chosen to at least achieve the ZF rates.

\section{Conclusions}
\label{sec:conclu}

In this paper, we proposed a simple line network model that captures the interference-limited behavior of a cellular downlink system.
Base station cooperation allows a joint encoding of user signals that can overcome the interference limitation; however, the capacity-achieving dirty paper coding scheme in this case may be too complex for practical implementation.
When all base stations in the network cooperate, zero-forcing is a simple linear precoding technique that offers good performance relative to DPC\@.
We proposed a new linear precoding technique called soft interference nulling (SIN) that performs better than or equal to ZF under full network coordination.
SIN precoding can be applied to maximize any general concave utility function of the user rates by solving a convex optimization problem, and the formulation is extended to the case when the terminals have multiple antennas.
Moreover, it is shown that SIN precoding with a limited coordination cluster size can outperform ZF with full network coordination at moderate SNRs.

\section*{Acknowledgment}
The authors would like to thank Dennis R. Morgan for helpful discussions.

\bibliographystyle{IEEEtran}
\bibliography{IEEEabrv,wrlscomm}

\end{document}